\documentclass{Interspeech}

\interspeechcameraready

\title{A Lightweight Hybrid Dual Channel Speech Enhancement System under Low-SNR Conditions}

\author[affiliation={1,2}]{Zheng}{Wang}
\author[affiliation={1,2}]{Xiaobin}{Rong}
\author[affiliation={3}]{Yu}{Sun}
\author[affiliation={1,2}]{Tianchi}{Sun}
\author[affiliation={1,2}]{Zhibin}{Lin}
\author[affiliation={1,2}]{Jing}{Lu}

\affiliation{Key Laboratory of Modern Acoustics}{Nanjing University}{China}
\affiliation{NJU-Horizon Intelligent Audio Lab}{Horizon Robotics}{China}
\affiliation{R\&D Centre}{Samsung Electronics (China)}{China}
\email{\{zheng.wang, xiaobin.rong\}@smail.nju.edu.cn, yuyu.sun@samsung.com, tianchi.sun@smail.nju.edu.cn,  \{zblin, lujing\}@nju.edu.cn}

\usepackage{comment}

\begin{document}

\maketitle

\begin{abstract}
    
    Although deep learning based multi-channel speech enhancement has achieved significant advancements, its practical deployment is often limited by constrained computational resources, particularly in low signal-to-noise ratio (SNR) conditions. In this paper, we propose a lightweight hybrid dual-channel speech enhancement system that combines independent vector analysis (IVA) with a modified version of the dual-channel grouped temporal convolutional recurrent network (GTCRN). IVA functions as a coarse estimator, providing auxiliary information for both speech and noise, while the modified GTCRN further refines the speech quality. We investigate several modifications to ensure the comprehensive utilization of both original and auxiliary information. Experimental results demonstrate the effectiveness of the proposed system, achieving enhanced speech with minimal parameters and low computational complexity.
    
\end{abstract}

\section{Introduction}

Speech enhancement aims to extract desired speech signals degraded by noise and interference. It serves as a crucial front-end module in applications, including human-machine interaction, video conferencing \cite{tan2019real}, and hearing aids \cite{li2020smart,modhave2016design}. With the rapid advancement of deep neural networks (DNNs), data-driven speech enhancement systems have demonstrated remarkable performance, surpassing traditional rule-based signal processing methods, which often struggle to maintain robustness in complex acoustic environments. 

In low-SNR conditions, where the target speech is overwhelmingly obscured by interference and noise \cite{hao2020unetgan}, speech enhancement techniques often fail to retain speech components. One approach to address this issue is the two-stage strategy \cite{hao2020masking}, which involves binary masking and spectrogram inpainting. The SNR-progressive model \cite{hou2024snr}, which integrates reliable pitch estimation and harmonic compensation \cite{le2023harmonic}, has also shown remarkable performance. However, the high computational demands of neural networks make them unsuitable for deployment on edge devices for real-time applications.

DNN-based multi-channel speech enhancement approaches have made great progress in recent years. One promising strategy is the neural beamformer, which preserves the structure of conventional beamformers, such as minimum variance distortionless response (MVDR) \cite{zhang2021adl}, by leveraging the ability of neural networks to effectively estimate the spatial covariance matrices (SCMs) of both speech and noise. Despite these advancements, the number of microphones required for these methods is often too large to enable practical deployment.

In response to the need for practical applications, recent works have focused on developing lightweight models that maintain competitive performance with reduced computational resources. For example, RNNoise \cite{valin2018hybrid} first performs at a low resolution and then utilizes a pitch filter for finer enhancement. GTCRN \cite{rong2024gtcrn}, which simplifies DPCRN \cite{le2021dpcrn} using multiple strategies, employs an equivalent rectangular bandwidth (ERB) filter bank to reduce feature redundancy, and uses grouped convolution \cite{ma2018shufflenet} and grouped RNN \cite{gao2018efficient} to shrink the model size. However, their speech enhancement performance in high noise level scenarios remains questionable.


\begin{figure}[tbp]
  \centering
  \includegraphics[width=0.8\linewidth]{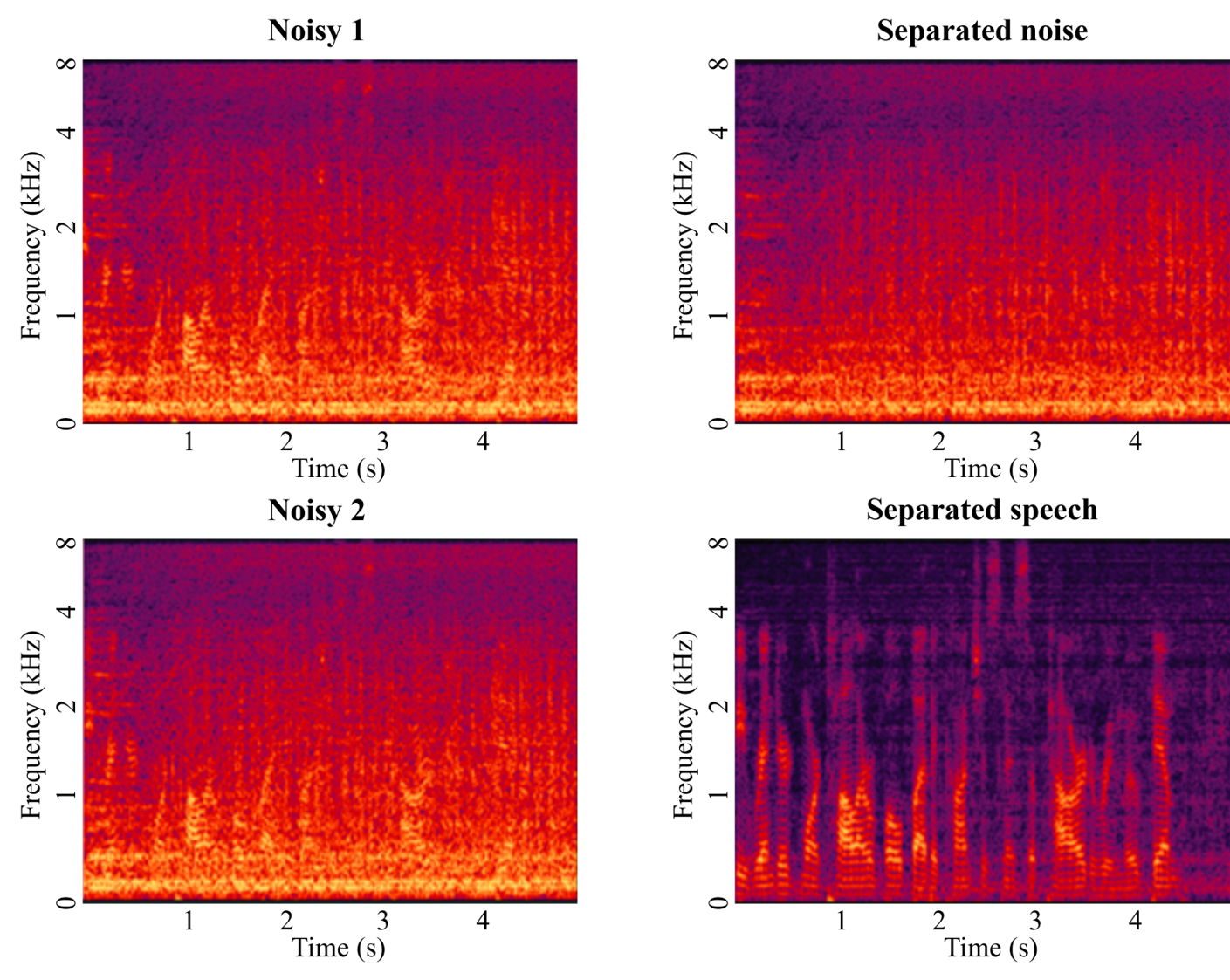}
  \caption{An example of IVA separation performance under low SNR conditions: left shows the dual-channel noisy spectrograms, and right shows the separated signal spectrograms.}
  \label{hybrid}
\end{figure}

\begin{figure*}[t]
        \begin{minipage}[b]{0.87\linewidth}
		\centering
		\subfloat[]{\includegraphics[width=\linewidth]{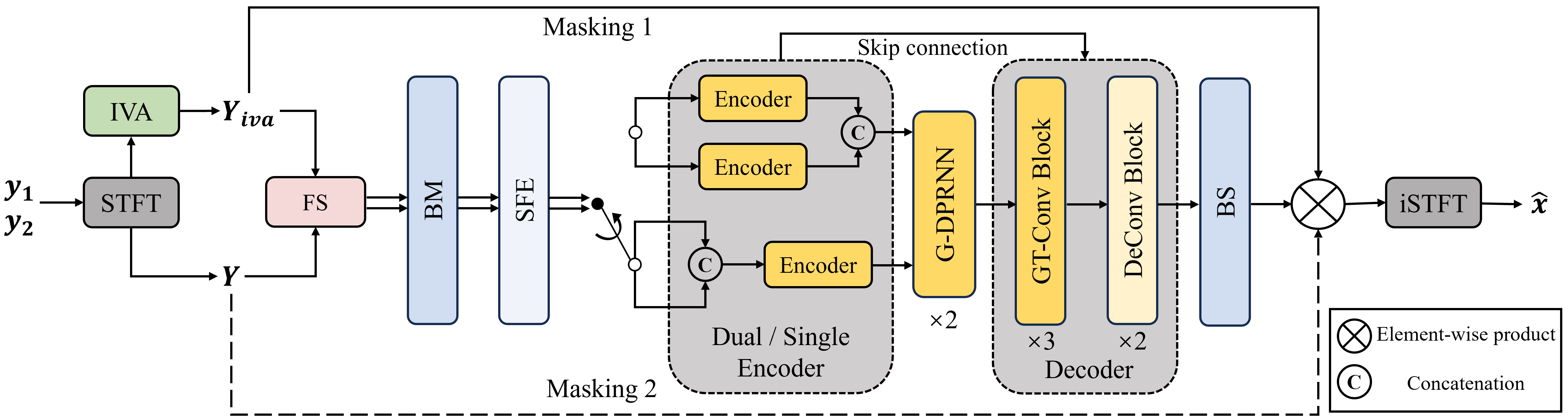}\label{model}}
	\end{minipage}
	\begin{minipage}[b]{0.11\linewidth}
		\centering
		\subfloat[]{\includegraphics[width=0.7\linewidth]{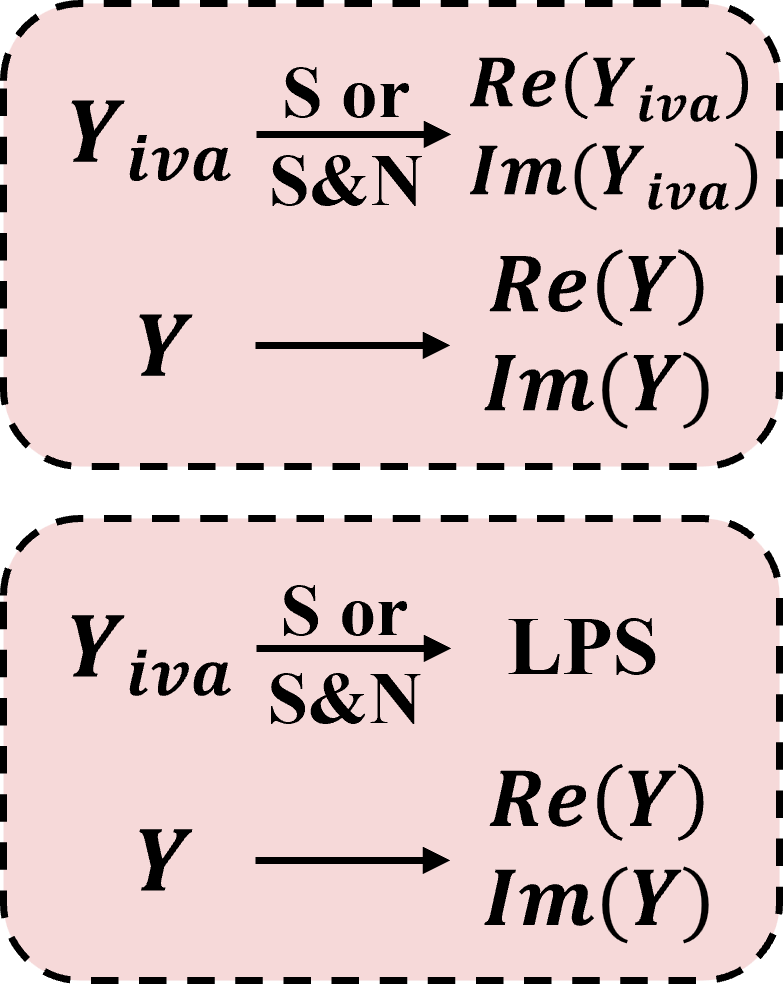}\label{feature}}
		\\
		\subfloat[]{\includegraphics[width=0.75\linewidth]{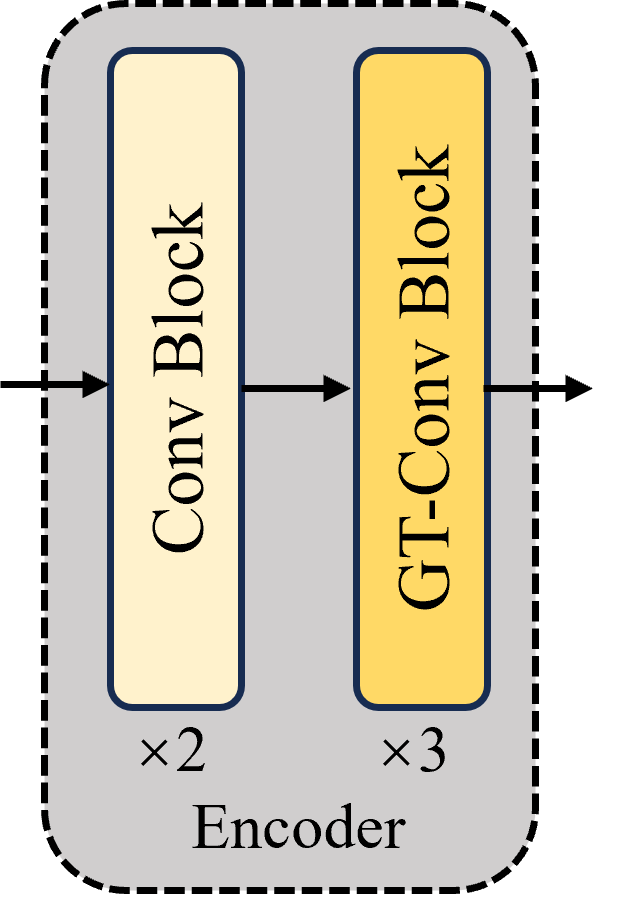}\label{encoder}}
	\end{minipage}
	\caption{(a) The framework of our proposed system, (b) two types of the feature selection module, where ``S or S\&N" refers to using either the speech channel alone or two separated channels, $Re(\cdot)$ and $Im(\cdot)$ represent the real and imaginary parts of the complex spectrogram, respectively, and LPS stands for log-power spectrogram, (c) the encoder module}
\end{figure*}

Independent vector analysis (IVA) \cite{kim2006independent, kim2006blind} is a traditional blind source separation method that uses statistical independence and spatial diversity to extract target signals with minimal computational cost compared to DNNs. While DNN-based methods often outperform IVA, the latter remains effective in separating speech and noise while preserving important features like harmonic structures, even in low-SNR conditions \cite{ruan2024speech}. An example in Figure \ref{hybrid} illustrates its effectiveness.

In this paper, we propose a lightweight hybrid dual-channel speech enhancement system that combines IVA with a modified dual-channel GTCRN to achieve effective speech enhancement in low-SNR conditions. IVA is employed as a coarse estimator to obtain preliminary separated speech and noise signals, which are then incorporated as auxiliary information to aid in speech preservation and interference suppression. Subsequently, our modified GTCRN refines these estimates, effectively integrating both the original noisy mixture and the separated source information provided by IVA to further enhance speech quality. We investigate multiple strategies, including the selection of input features, auxiliary information, and masking approaches, along with the adoption of a dual-encoder structure, to ensure the comprehensive utilization of diverse information. Experimental results validate the effectiveness of our proposed hybrid system, demonstrating its ability to achieve competitive performance with small-scale parameters and low computational complexity, making it well-suited for real-time applications.

\section{Problem formulation}

We consider the task of extracting the target speech signal from ambient noise and other interference. The problem in the time domain can be expressed as follows:
\begin{gather}
    x_m(t)=h_m(t)*s(t),\\
    y_m(t)=x_m(t)+n_m(t),
\end{gather}
where $t$ is the time index, $s(t)$ represents the non-reverberant speech signal, $h_m(t)$ represents the room impulse response (RIR) between the speaker and the $m$th microphone, $*$ denotes the convolution operation, and $y_m(t)$, $x_m(t)$, $n_m(t)$ respectively represent the mixture, the speech, and the noise signals captured by the $m$th microphone. It can be further converted into the short-time Fourier transform (STFT) domain as 
\begin{equation}
    Y_m(k,l)=H_m(k,l)S(k,l)+N_m(k,l),
\end{equation}
where $k$ is the frequency index, $l$ is the frame index, $Y_m(k,l)$, $H_m(k,l)$, $S(k,l)$, and $N_m(k,l)$ are the STFT-domain representations of $y_m(t)$, $h_m(t)$, $s(t)$ and $n_m(t)$, respectively.

Our goal is to develop a lightweight dual-channel speech enhancement system $f$ to obtain the enhanced signal expressed as:
\begin{gather}
    \hat{x}(t)=f(y_1(t),y_2(t)),
\end{gather}
where $\hat{x}(t)$ represents the enhanced signal. 

\section{Proposed system}

As depicted in Figure \ref{model}, our proposed system is built upon the GTCRN architecture. We introduce the dual-channel version of the model and incorporate several modifications to ensure the comprehensive utilization of diverse information. These modifications include feature selection, auxiliary information selection, masking approach selection, and the adoption of a dual-encoder structure. The system consists of several key components: a feature selection (FS) module, a band merging (BM) module, a subband feature extraction (SFE) module, an encoder module, a grouped dual-path RNN (G-DPRNN) module, a decoder module, and a band splitting (BS) module. A skip connection is employed to mitigate potential information loss during the encoding process. After the band splitting step, the outputs of decoder, interpreted as the real and imaginary parts of the complex ratio mask (CRM) \cite{williamson2015complex}, are multiplied with either the IVA output (Masking 1) or the original noisy input (Masking 2) to yield the estimated speech.

\subsection{Model input}

We obtain the complex spectrogram $Y \in \mathbb{C}^{C\times T \times F}$ of the noisy mixture by applying STFT, where $C$, $T$ and $F$ denote the channel, time, and frequency dimensions, respectively. The coarse estimation of sources, $Y_{iva} \in \mathbb{C}^{C\times T \times F}$, is obtained by IVA. As illustrated in Figure \ref{feature}, two distinct types of IVA features are available for selection: the complex spectrogram and the log-power spectrogram (LPS). Additionally, the potential benefits of utilizing separated speech and noise information is another critical issue to be investigated. The real and imaginary parts of the original noisy spectrogram, along with the selected IVA feature, are concatenated along the channel dimension. To downsample the spectrogram features, a BM operation is first applied, which can later be inverted to the original resolution using a BS operation. The SFE module begins by merging each frequency band with adjacent bands, followed by stacking each subband unit along the channel dimension via a reshape operation. This process extends the subband relationships from the frequency dimension to the channel dimension, enabling the subsequent convolutional layers to leverage frequency information more efficiently.

\subsection{Aux-IVA}
We adopt Aux-IVA, a fast and stable IVA algorithm that utilizes the auxiliary-function technique \cite{ono2011stable,taniguchi2014auxiliary}, in the IVA module. Given the complex spectrogram $\mathbf{Y}(k)=[Y_1(k), Y_2(k)]^T$, the demixing matrices $W=[\mathbf{w_1}(k),\mathbf{w}_2(k)]^H$, initially set to an identity matrix for each frequency bin, are iteratively updated to obtain the separated complex spectrogram. The update rules for the auxiliary variables and the demixing matrices are as follows:
\begin{equation}
    r_m=\sqrt{\sum_{k=1}^F\left\vert\mathbf{w}_m^H(k)\mathbf{Y}(k)\right\vert^2},
\end{equation}
\begin{equation}
    V_m(k)=\mathbb{E}[\frac{G'(r_m)}{r_m}\mathbf{Y}(k)\mathbf{Y}^H(k)],
\end{equation}
\begin{equation}
    \mathbf{w}_m(k)=(W(k)V_m(k))^{-1}\mathbf{e}_k,
\end{equation}
\begin{equation}
    \mathbf{w}_m(k)=\mathbf{w}_m(k)/\sqrt{\mathbf{w}_m^H(k)V_m(k)\mathbf{w}_m(k)},
\end{equation}
where $m=1,2$ represents the microphone index, $G(\cdot)$ is a contrast function, $\mathbb{E}[\cdot]$ denotes the expectation operator. 
\subsection{Encoder and decoder}

We explore two types of encoder frameworks: single encoder and dual encoder \cite{chidambaram2018learning}. As depicted in Figure \ref{encoder}, every encoder consists of two convolution (Conv) blocks, with each block consisting of a convolution layer, a batch normalization, and a PReLU activation, along with three grouped temporal convolution (GT-Conv) blocks \cite{rong2024gtcrn}. The GT-Conv blocks enhance long-range temporal dependency modeling by incorporating temporal dilation into depth-wise convolutions. They split the input features into two branches, one of which undergoes a sequence of two 2D point-wise convolutions (P-Conv2D) layers and a 2D dilated depth-wise convolution (DD-Conv2D) layer. The outputs of both branches are then concatenated, and a channel shuffle operation facilitates information exchange between them. The decoder mirrors the encoder's structure, with the Conv blocks replaced by DeConv blocks, where the convolution layers are substituted with transposed convolution layers. Additionally, the activation of the last DeConv block is replaced by tanh to constrain output values within the range of $(-1,1)$.

\subsection{Grouped DPRNN}

The combined grouped RNN (GRNN) and DPRNN \cite{luo2020dual} architecture, referred to as G-DPRNN \cite{rong2024gtcrn}, splits the input features into two disjoint groups. Each group is then processed by a recurrent layer, followed by a representation rearrangement layer to obtain output features. For intra-frame modeling, we utilize grouped bidirectional GRUs to capture the spectral patterns within a single frame. For inter-frame modeling, we employ grouped unidirectional GRUs to exploit the temporal dependencies within a specific frequency bin.

\subsection{Loss function}
Our model is trained on the hybrid loss function, which consists of scale-invariant signal-to-noise ratio (SISNR) loss and complex compressed mean-squared error (ccMSE) loss \cite{braun2021consolidated}. The overall loss function is given by
\begin{align}
\mathcal{L} &= \alpha \mathcal{L}_{\text{SISNR}}(\hat{x}, x) + (1 - \beta) \mathcal{L}_{\text{mag}}(\hat{X}, X) \nonumber \\
&\quad + \beta \left( \mathcal{L}_{\text{real}}(\hat{X}, X) + \mathcal{L}_{\text{imag}}(\hat{X}, X) \right),
\end{align}
where $\hat{x}$ and $x$ denote the enhanced and target waveform, $\hat{X}$ and $X$ denote the complex spectrogram of the enhanced and target, respectively. $\alpha$ and $\beta$ are the weighting factors, which are respectively set to 0.01 and 0.3 in this work. 

Each term in the aforementioned formula is calculated as follows:


\begin{equation}
    \mathcal{L}_{\text{SISNR}} = -\log_{10} (\frac{\lVert \hat{x}_t \rVert^2}{\lVert \hat{x} - {x}_t \rVert^2});\hat{x}_t =  \frac{\langle \hat{x}, x \rangle x}{\lVert x \rVert^2},
\end{equation}
\begin{equation}
    \mathcal{L}_{\text{mag}}(\hat{X}, X) = \text{MSE}(|\hat{X}|^{0.3}, |X|^{0.3}),
\end{equation}
\begin{equation}
    \mathcal{L}_{\text{real}}(\hat{X}, X) = \text{MSE}(\hat{X}_r / |\hat{X}|^{0.7}, X_r / |X|^{0.7}),
\end{equation}
\begin{equation}
    \mathcal{L}_{\text{imag}}(\hat{X}, X) = \text{MSE}(\hat{X}_i / |\hat{X}|^{0.7}, X_i / |X|^{0.7}).
\end{equation}

\section{Experiment}


\subsection{Dataset}
We generate the simulated dataset with the image method \cite{allen1979image}, with dual-channel RIRs based on a linear array with two microphones placed \SI{4}{cm} apart. The room size ranges from \SI{3}{m}$\times$\SI{3}{m}$\times$\SI{2.5}{m} to \SI{10}{m}$\times$\SI{10}{m}$\times$\SI{3}{m}, and the reverberation time ($\text{RT}_{60}$) ranges from \SI{0.1}{s} to \SI{0.4}{s}. The distance from the source to the array is randomly selected from \{\SI{0.5}{m}, \SI{1}{m}, \SI{2}{m}, \SI{3}{m}\}, with the direction of arrival (DOA) difference between the target speech and interference noise being greater than \SI{5}{°}. We convolve the speech dataset from the DNS-3 challenge \cite{reddy2021icassp} with these dual-channel RIRs to generate the simulated speech signals. The early reflection (\SI{50}{ms}) of the first channel is preserved as the training target. For the noise signals, we select data from the DNS-3 and DCASE \cite{dohi2022description} datasets. During training, the SNR ranges from \SI{-10}{dB} to \SI{0}{dB}. The validation set maintains this range of SNR, with 500 noisy-clean pairs in each case. For the test set, we set three SNR levels: {\SI{-12.5}{dB}, \SI{-7.5}{dB}, and \SI{-2.5}{dB}}, each with 500 noisy-clean pairs. All utterances are sampled at \SI{16}{kHz}.


\subsection{Implementation details}
The STFT is performed using a square root Hanning window of a length of 512 (\SI{32}{ms}) and a hop length of 256 (\SI{16}{ms}). In the BM module, 192 high-frequency bands are mapped to 64 ERB bands, while 65 low-frequency bands keep unaltered. For the SFE module, a kernel size of 3 is used. The two Conv blocks share a common output channel number of 16 for the single encoder case and 12 for the dual encoder case, ensuring comparable computational complexity for a fair comparison. The kernel size is set to (1,5) with a stride of (1,2), while the second layer uses a group size of 2 to reduce computational complexity. The DD-Conv2D layers in the three GT-Conv blocks have a common channel number of 16, with a kernel size of (3,3), and time dilation values of 1,2 and 5, respectively.

The models are trained using the Adam optimizer \cite{kingma2014adam} with a warm-up phase. The linear-warm-up-cosine-annealing-learning-rate scheduler is employed, where the learning rate increases linearly during the initial training phase and decreases following a cosine function. In our training, the batch size is set to 8, and the number of steps for each epoch is 1250. The number of warm-up steps and total steps are set to 25000 (20 epochs) and 250000 (200 epochs), respectively. The minimum and maximum learning rates are set to $10^{-6}$ and $10^{-3}$.

\begin{table*}[t]
    \scriptsize
    \centering
    \caption{Results of the ablation study on the simulated validation set.  \textbf{BOLD} indicates the best score in each metric.}
    \vspace{-8pt}
    \label{tab:ablation}
    \resizebox{\linewidth}{!}{
    \begin{tabular}{cccccccccc}
        \toprule
        \multirow{2}{*}{Metrics} &  &   & & \multirow{2}{*}{PESQ} & \multirow{2}{*}{STOI($\times$100)} & \multirow{2}{*}{DNSMOS-P.808} & \multicolumn{3}{c}{DNSMOS-P.835}\\
         &  &   &  &  &  & & SIG & BAK & OVRL \\
        \midrule
        IDs & Methods & Para. (k)
        & MACs (M/s) &  &  &  &  &  &  \\
        \midrule
        - & Noisy & - & - & 1.08 & 55.48 & 2.19 & 1.21 & 1.16 & 1.10 \\
        \midrule
        1 & IVA-S+Complex Feature+Masking 1 & 24.39 & 43.20 & 1.46 & 77.83 & 3.01 & 2.37 & 3.76 & 2.11 \\
        2 & IVA-S+Complex Feature+Masking 2 & 24.39 & 43.20 & 1.51 & 79.13 & 3.17 & 2.59 & 3.77 & 2.28\\
        3 & IVA-S+LPS Feature+Masking 1 & 24.15 & 41.44 & 1.49 & 78.64 & 3.11 & 2.35 & 3.76 & 2.09\\
        4 & IVA-S\&N+LPS Feature+Masking 1 & 24.39 & 43.20 & 1.53 & 79.63 & 3.11 & 2.37 & 3.76 & 2.11 \\
        5 & IVA-S+LPS Feature+Masking 2 & 24.15 & 41.44 & 1.57 & 79.11 & 3.25 & \textbf{2.61} & 3.81 & 2.31 \\
        6 & IVA-S\&N+LPS Feature+Masking 2 & 24.39 & 43.20 & \textbf{1.61} & 79.57 & \textbf{3.29} & 2.60 & \textbf{3.83} & \textbf{2.32}\\
        7 & IVA-S\&N+LPS Feature+Masking 2+Dual-Encoder & 25.57 & 45.63 & 1.60 & \textbf{79.99} & 3.26 & 2.56 & 3.78 & 2.28 \\
        \bottomrule
    \end{tabular}
    }
\end{table*}


\begin{table*}[t]
    \tiny
    \centering
    \caption{Results comparison with baselines on the simulated test set. \textbf{BOLD} indicates the best score in each metric.}
    \vspace{-8pt}
    \label{tab:se_performance}
    \setlength{\tabcolsep}{5.4pt}
    \begin{tabular}{ccccccccccccccccccccc}
        \toprule
        {Metrics} &  &  & \multicolumn{3}{c}{PESQ} & \multicolumn{3}{c}{STOI($\times$100)} & \multicolumn{3}{c}{DNSMOS-P.808} & \multicolumn{3}{c}{DNSMOS-SIG}& \multicolumn{3}{c}{DNSMOS-BAK}& \multicolumn{3}{c}{DNSMOS-OVRL}\\
        \cmidrule(r){4-6} \cmidrule(r){7-9} \cmidrule(r){10-12} \cmidrule(r){13-15}\cmidrule(r){16-18}\cmidrule(r){19-21} SNR (dB) & Para. (k)
        & MACs (M/s) & -12.5 & -7.5 & -2.5 & -12.5 & -7.5 & -2.5 & -12.5 & -7.5 & -2.5 & -12.5 & -7.5 & -2.5 & -12.5 & -7.5 & -2.5 & -12.5 & -7.5 & -2.5 \\
        \midrule
        Noisy & - & - & 1.05 & 1.05 & 1.06 & 42.00 & 50.46 & 61.07 & 2.17 & 2.17 & 2.20 & 1.17 & 1.19 & 1.25 & 1.14 & 1.14 & 1.18 & 1.08 & 1.09 & 1.12\\
        Aux-IVA & - & - & 1.07 & 1.13 & 1.20 & 62.52 & 70.22 & 73.37 & 2.36 & 2.53 & 2.66 & 1.61 & 2.00 & 2.24 & 1.40 & 1.67 & 1.93 & 1.31 & 1.53 & 1.72\\
        GTCRN & 23.43 & 32.07 & 1.09 & 1.16 & 1.31 & 47.46 & 61.12 & 72.94 & 2.39 & 2.60 & 2.90 & 1.73 & 1.99 & 2.36 & 3.43 & 3.65 & 3.79 & 1.49 & 1.75 & 2.09\\
        DC-GTCRN & 23.91 & 35.59 & 1.15 & 1.29 & 1.50 & 60.21 & 71.27 & 79.69 & 2.57 & 2.81 & 3.09 & 1.82 & 2.14 & 2.50 & 3.56 & 3.71 & 3.82 & 1.61 & 1.90 & 2.22\\
        DC-GTCRN-L & 34.80 & 49.10 & 1.17 & 1.32 & 1.54 & 61.16 & 72.13 & 80.39 & 2.59 & 2.86 & 3.14 & 1.88 & 2.21 & 2.56 & 3.60 & 3.75 & 3.85 & 1.66 & 1.96 & 2.28\\
        \textbf{Proposed} & 24.39 & 43.20 & \textbf{1.39} & \textbf{1.58} & \textbf{1.71} & \textbf{72.38} & \textbf{79.01} & \textbf{81.96} & \textbf{3.03} & \textbf{3.26} & \textbf{3.39} & \textbf{2.36} & \textbf{2.62} & \textbf{2.74} & \textbf{3.76} & \textbf{3.85} & \textbf{3.87} & \textbf{2.09} & \textbf{2.34} & \textbf{2.44}\\
        \bottomrule
    \end{tabular}
\end{table*}


\section{Results}

\subsection{Evaluation metrics}
The evaluation is conducted using the objective metrics, including perceptual evaluation of speech quality (PESQ) \cite{rix2001perceptual} and short-time objective intelligibility (STOI) \cite{taal2010short}. Additionally, DNN-based non-intrusive subjective metrics DNSMOS P.808 \cite{reddy2021dnsmos} and DNSMOS P.835 \cite{reddy2022dnsmos} are also employed.

\subsection{Ablation study}
We conduct an ablation study on our modified GTCRN to evaluate the impact of various factors, including the use of speech and noise information from IVA, the type of feature, the type of masking approach, and the adoption of the dual-encoder, as shown in Table \ref{tab:ablation}. We compare the performance of different masking approaches, as seen in IDs 1 and 2, IDs 3 and 5, and IDs 4 and 6. It is clear that Masking 2 (IDs 2, 5, 6) outperforms Masking 1 (IDs 1, 3, 4), which can be attributed to the coarse estimation and speech preservation provided by IVA. A comparison between ID-1 and ID-3 highlights the effectiveness of the LPS feature, which improves all metrics, indicating that phase information is not as crucial. Compared to ID-5, the inclusion of the noise information provided by IVA (ID-6) leads to substantial improvements in nearly all metrics, despite a slight degradation in SIG is observed. This is because the two separated signals allow for a more comprehensive capture of the noisy mixture, while the speech channel focuses solely on the speech component. Finally, the results for ID-7 exhibit noticeable declines across most metrics, despite a marginal improvement in STOI over ID-6. We attribute this to the computational limitations of our lightweight network, which restrict the ability to fully leverage the potential benefits of the dual-encoder framework. Consequently, we select the best-performing method in the table (ID-6) for comparison with the baseline models.

\subsection{Results comparison with baselines}



Four methods are selected as baselines: (a) Aux-IVA, (b) GTCRN, (c) DC-GTCRN, a dual-channel version of GTCRN, and (d) DC-GTCRN-L, a larger-scale version of DC-GTCRN. The results of the simulated test set are presented in Table \ref{tab:se_performance}. Compared to the noisy mixture, the output of Aux-IVA shows a significant improvement in the STOI metric, demonstrating its effectiveness in preserving speech while suppressing noise. By incorporating the second channel input, DC-GTCRN leverages additional inter-channel information, such as phase differences, leading to significant improvements across all evaluation metrics compared to GTCRN, which shows limited performance in low SNR conditions. Although DC-GTCRN-L achieves improved performance with an increased number of parameters, the performance gap between this baseline and our proposed method remains substantial, even with the additional computational overhead. By integrating Aux-IVA, our proposed method attains the highest scores, further validating the efficacy of auxiliary information and underscoring the superiority of our hybrid approach. The parameters and computational loads reported in the table include the Aux-IVA module, which contributes a negligible increase in parameters and only 0.20 MMACs per second per iteration.


\begin{figure}[htbp]
\centering
\begin{minipage}[b]{0.32\linewidth}
  \centering
  \includegraphics[width=\linewidth]{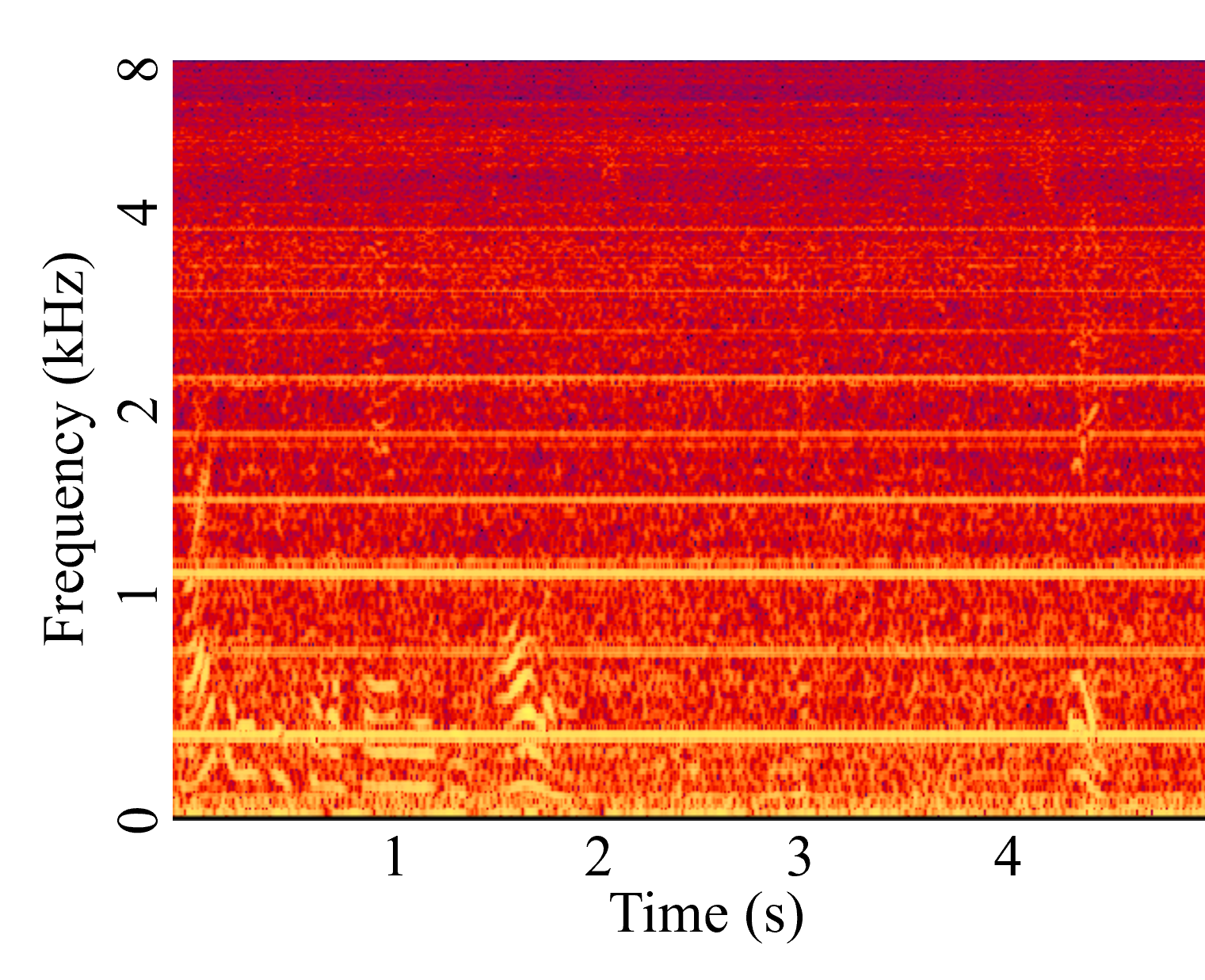} 
  \subcaption{Noisy}\label{fig:noisy}
\end{minipage}
\hfill
\begin{minipage}[b]{0.32\linewidth}
  \centering
  \includegraphics[width=\linewidth]{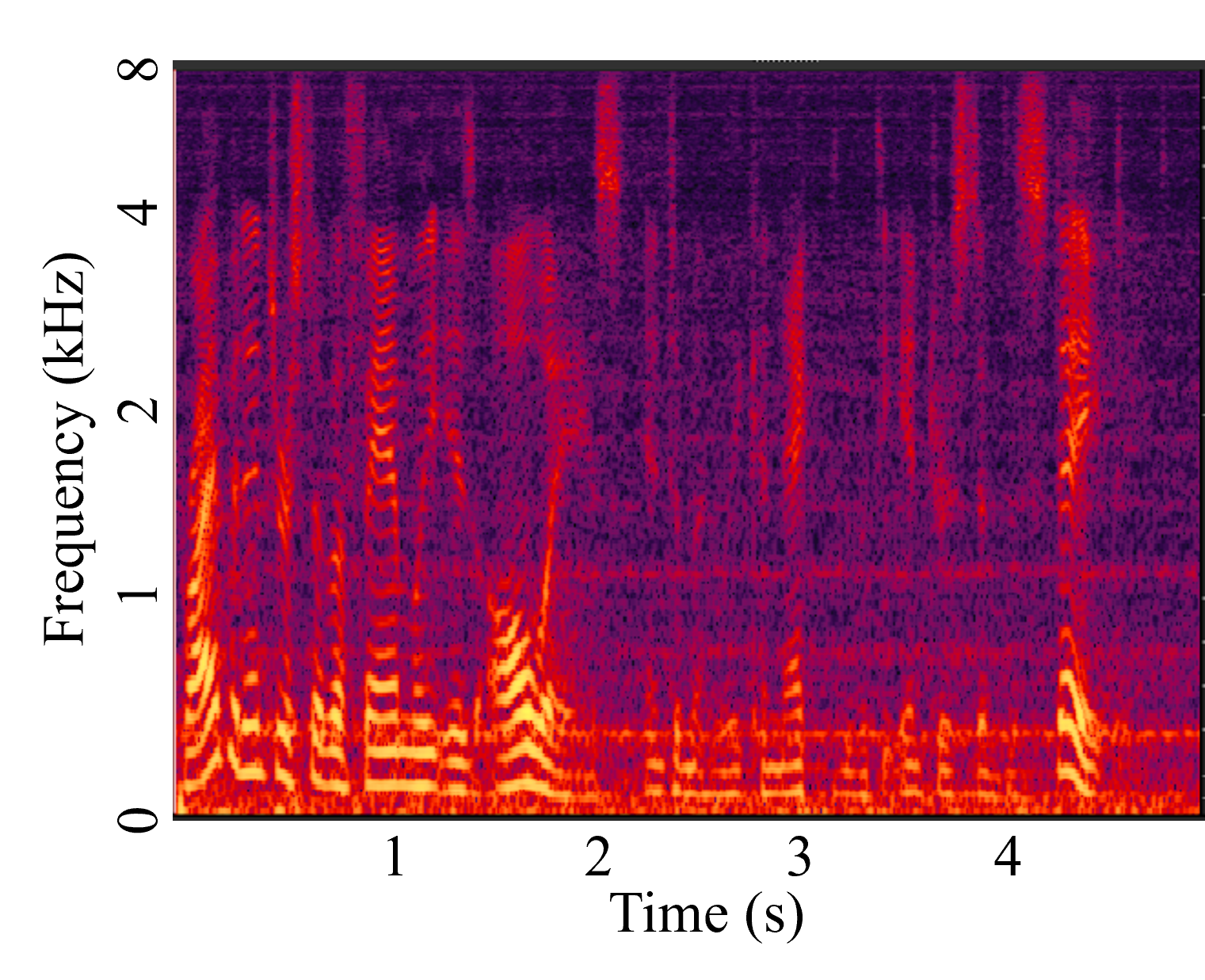} 
  \subcaption{Aux-IVA}\label{fig:iva}
\end{minipage}
\hfill
\begin{minipage}[b]{0.32\linewidth}
  \centering
  \includegraphics[width=\linewidth]{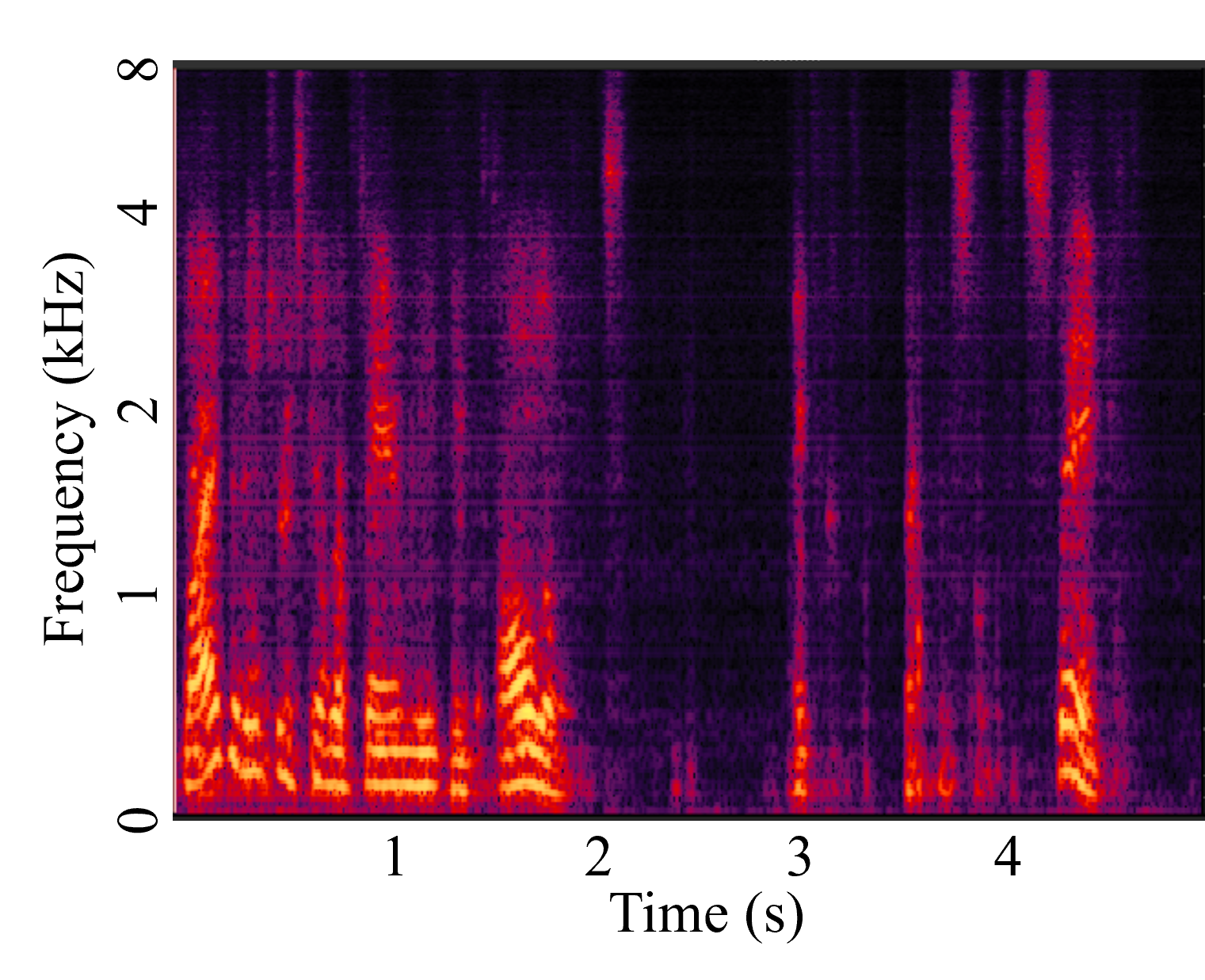} 
  \subcaption{GTCRN}\label{fig:1ch}
\end{minipage}
\hfill
\begin{minipage}[b]{0.32\linewidth}
  \centering
  \includegraphics[width=\linewidth]{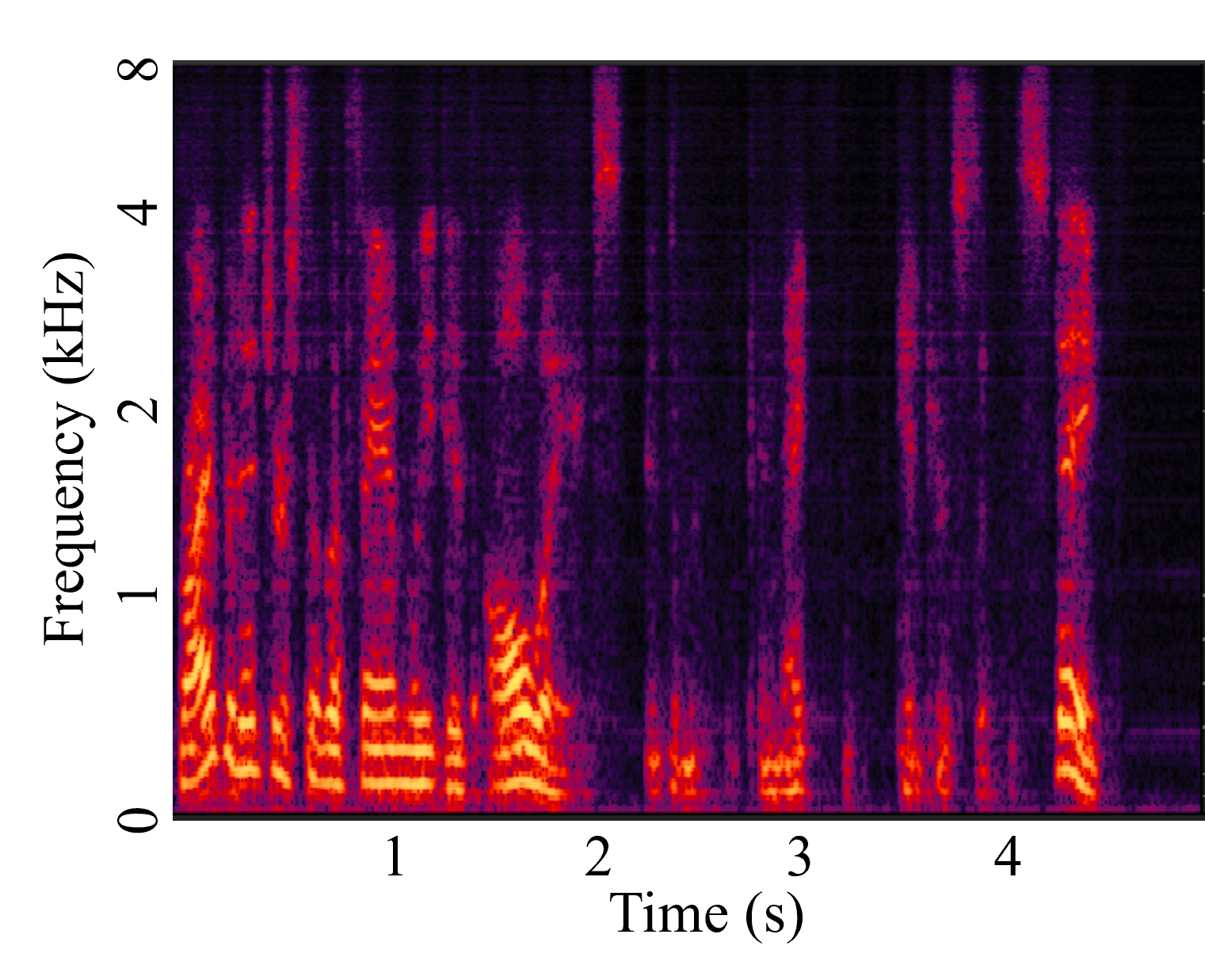} 
  \subcaption{DC-GTCRN}\label{fig:2ch}
\end{minipage}
\hfill
\begin{minipage}[b]{0.32\linewidth}
  \centering
  \includegraphics[width=\linewidth]{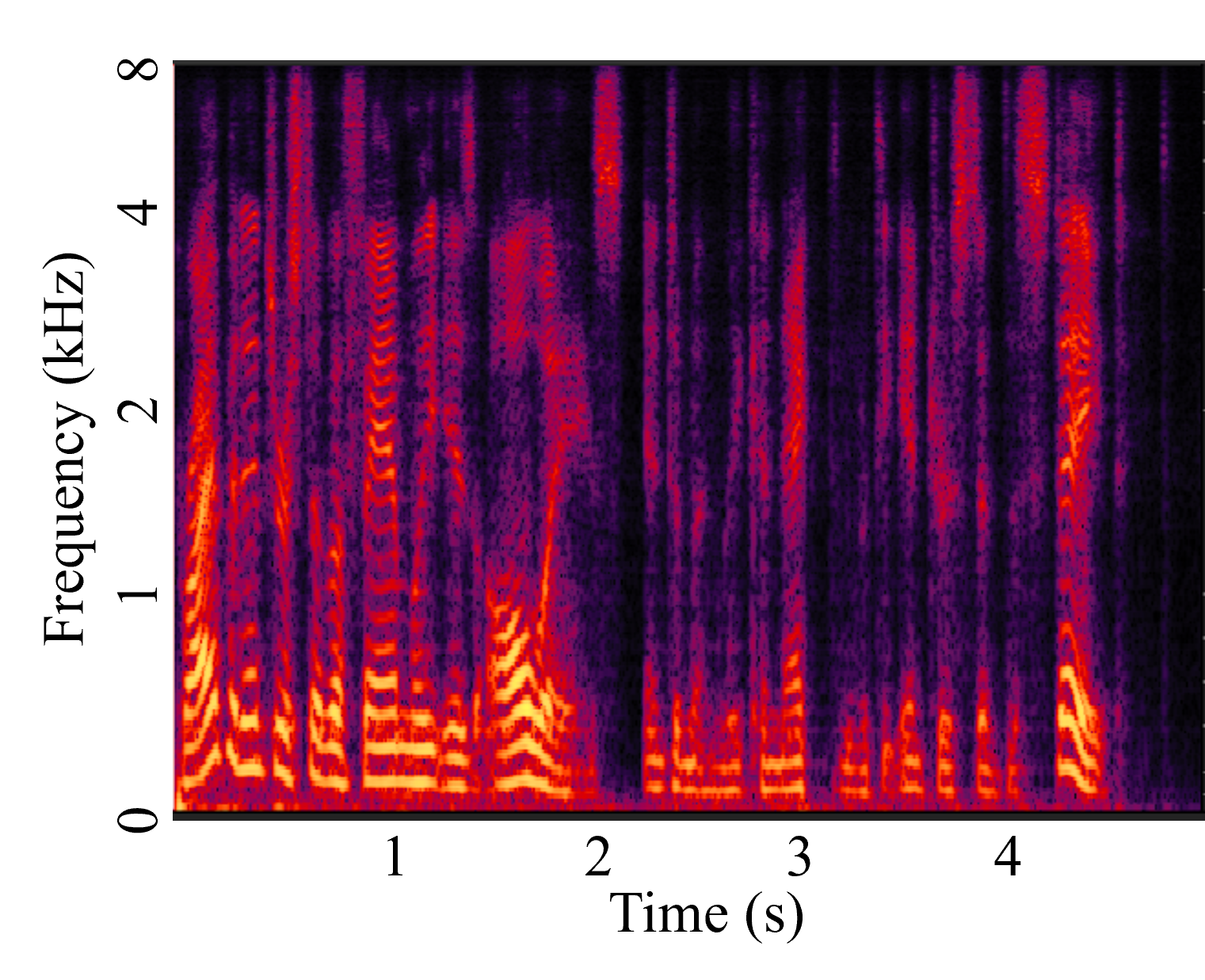} 
  \subcaption{Proposed}\label{fig:gtcrn}
\end{minipage}
\hfill
\begin{minipage}[b]{0.32\linewidth}
  \centering
  \includegraphics[width=\linewidth]{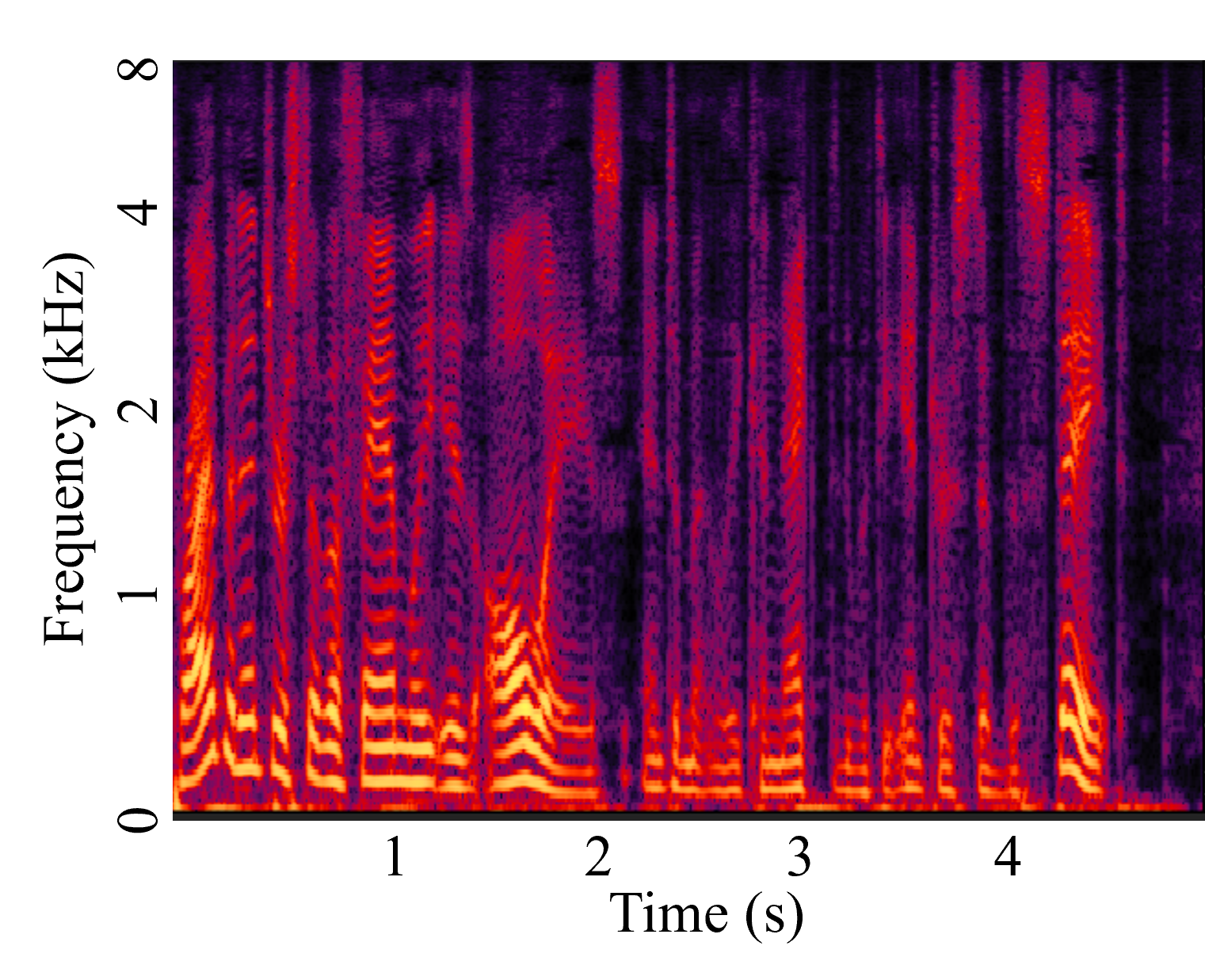} 
  \subcaption{Clean}\label{fig:clean}
\end{minipage}

\caption{Typical spectrograms}
\label{fig:spectrograms}
\end{figure}

A set of typical audio samples is presented in Figure \ref{fig:spectrograms}, clearly demonstrating that our proposed method excels in both speech preservation and noise suppression. The original noisy spectrogram is predominantly dominated by noise, making it difficult to identify speech components. It is evident that the baseline models fail to effectively enhance speech under such low SNR conditions. In contrast, the Aux-IVA result extracts speech components despite some residual noise, while our method’s enhanced result more effectively suppresses noise, and retains more speech details, including harmonic components, further highlighting its effectiveness. Audio samples are available: \url{https://github.com/Max1Wz/H-GTCRN}.

\section{Conclusion}
In this paper, we propose a hybrid dual-channel speech enhancement system designed for low-SNR conditions, integrating IVA and a modified GTCRN. Aux-IVA acts as a coarse estimator, providing auxiliary information, while the GTCRN further refines the speech quality. Through various architecture modifications, both the original and auxiliary information are fully leveraged. With only a minimal increase in parameters and computational complexity, the proposed system effectively enhances speech. Experimental results validate its effectiveness.


\section{Acknowledgements}

This work is supported by the National Natural Science Foundation of China (Grant No. 12274221) and the AI \& AI for Science Project of Nanjing University.


\bibliographystyle{IEEEtran}
\bibliography{mybib}

\end{document}